\documentclass[12pt,preprint]{aastex}
\usepackage{amsmath}

\shorttitle{Jitter Polarization and Depolarization of GRBs}

\shortauthors{Mao et al.}

\begin{document}


\title{Linear Polarization, Circular Polarization, and Depolarization of Gamma-ray Bursts: A Simple Case of Jitter Radiation}


\author{
Jirong Mao\altaffilmark{1,2,3} and
Jiancheng Wang\altaffilmark{1,2,3}
}
\altaffiltext{1}{Yunnan Observatories, Chinese Academy of Sciences, 650011 Kunming, Yunnan Province, China}
\altaffiltext{2}{Center for Astronomical Mega-Science, Chinese Academy of Sciences, 20A Datun Road, Chaoyang District, Beijing, 100012, China}
\altaffiltext{3}{Key Laboratory for the Structure and Evolution of Celestial Objects, Chinese Academy of Sciences, 650011 Kunming, China}

\email{jirongmao@mail.ynao.ac.cn}

\begin{abstract}
Linear and circular polarizations of gamma-ray bursts (GRBs) have been detected during recent years. We adopt a simplified model to investigate GRB polarization characteristics in this paper. A compressed two-dimensional turbulent slab containing stochastic magnetic fields is considered, and jitter radiation can produce the linear polarization under this special magnetic field topology. Turbulent Faraday rotation measure (RM) of this slab makes strong wavelength-dependent depolarization.
The jitter photons can also scatter with those magnetic clumps inside the turbulent slab, and a nonzero variance of the Stokes parameter $V$ can be generated.
Furthermore, the linearly and circularly polarized photons in the optical and radio bands may suffer heavy absorptions from the slab. Thus we consider the polarized jitter radiation transfer processes.
Finally, we compare our model results with the optical detections of GRB 091018, GRB 121024A, and GRB 131030A.
We suggest simultaneous observations of GRB multi-wavelength polarization in the future.
\end{abstract}




\section{Introduction}
Polarization measurements of gamma-ray bursts (GRBs) have been performed during recent years. Linear polarization of GRB prompt emissions can be detected in the high-energy band, and large polarization degrees are given. GRB 021206 with $80\%$ polarization degree was obtained by {\it RHESSI} \citep{coburn03}. From the {\it INTEGRAL} observation, GRB 041219A polarization has 98\% degrees measured by \citet{kalemci07} and 63\% degrees measured by \citet{mcglynn07}.
The GRB polarimeter onboard {\it IKAROS} had the polarization detections for three GRBs. The polarization degree of GRB 100826A is $27\%$ \citep{yonetoku11}.
High polarization degrees of 70\% for GRB 110301A and 84\% for GRB 110721A were also reported \citep{yonetoku12}.
We note that GRBs also emit polarization photons in the optical band. The linear polarization upper limits of GRB 990123 (2.3\% degree), GRB 060418 (8\% degree), and GRB 110205A (16\% degree) were obtained \citep{hjorth99,mundell07,cucchi11}. The early-time linear polarizations of GRB 090102 (10\% degree), GRB 091208B (10.4\% degree), GRB 131030A (2\% degree), and GRB 150301B (8\% degree) were detected by \citet{steele09}, \citet{uehara12}, \citet{king14}, and \citet{gor16}, respectively. The late-time linear polarizations of GRB 990510 (1.7\% degree; see Covino et al. 1999 and Wijers et al. 1999) and GRB 020405 (9.9\% degree, see Bersier et al. 2003) were confirmed as well. Moreover, the temporal variability of the linear polarization was found in GRB 990712 \citep{rol00}, GRB 021004 \citep{rol03}, GRB 030329 \citep{greiner03}, GRB 091018 \citep{wier12}, and GRB 120308A \citep{mundell13}. Recently, both the circular polarization (0.6\% degree) and the linear polarization (4\% degree) of GRB 121024A were successfully achieved \citep{wier14}.

Synchrotron radiation, which is the emission of relativistic electrons in a large-scale magnetic field, is often applied as one popular mechanism to explain GRB polarization. The linear polarization degree is generally calculated by $\Pi_l=(3p_e+3)/(3p_e+7)$, where $p_e$ is the power-law index of a given electron energy spectrum \citep{westford59,rybicki79}. We obtain $\Pi_l=71\%$ when $p_e=2.2$. A more complicated consideration with GRB afterglow hydrodynamics was given by \citet{lan16}. The Compton scattering process may also produce the GRB polarization \citep{chang14}. The circular polarization can be intrinsically generated by the synchrotron mechanism. The degree is small, and it is proportional to $1/\gamma$, where $\gamma$ is the Lorentz factor of a relativistic electron (Legg \& Westfold 1968; Nava et al. 2015; However see de B\'{u}rca \& Sheare 2015). Several additional issues have been concerned. First, the high-degree linear polarization may be interpreted by the synchrotron radiation inside a structured GRB jet \citep{granot03,rossi04}. The jet structure was also applied to explain the low-degree polarization detected in the optical band \citep{ghisellini99}. Second, because of the Faraday rotation, the depolarization effect as a function of observational wavelength should be considered \citep{burn66}. One may expect that long-wavelength photons have the low-degree polarization. Third,
if the emission photons pass through a dense plasma environment, one may solve the radiation transfer equation to obtain the Stokes parameters \citep{sazonov68,sazonov69a,sazonov69b,jones77,matsumiya03,toma08}. Fourth, if we propose the emissions in some magnetic patches, the net linear polarization degree is $1/\sqrt{N}$ times of the original one, where $N$ is the patch number \citep{gruzinov99}; This leads us to consider other physical processes related to tangled magnetic fields.

Turbulent plasma can produce the Faraday rotation and make some depolarization effects. The stochastic Faraday rotation and the polarization features have been explored, and the bandwidth depolarization effect presented by $\Pi\propto exp(-\lambda^4)$ is extremely strong \citep{simonetti84,melrose93a,melrose93b,melrose98}. If the structure function of the Faraday RM has a turbulent cascade, the polarization turns to drop as $\Pi\propto \lambda^{-4/\zeta_p}$, where $\zeta_p$ is the index of a given turbulent energy spectrum \citep{tribble91}. Furthermore, some weak signals of circular polarization can be produced by this turbulent plasma screen. \citet{macquart00a} proposed a nonzero variance of the Stokes parameter $V$, and $\langle V^2\rangle \propto \langle I\rangle^2$ was given. The astrophysical scenario is the following \citep{macquart00b}: the turbulent plasma screen can be treated as a birefringence, and the random refractions make up the left-side and right-side circularly polarized shock fronts. The random pattern of one wave front is slightly different from the other, such that the circular polarization occurs.

The turbulent screen contains many random and small-scale magnetic elements, and relativistic electrons in this screen have emissions. Jitter radiation, which is the radiation of the relativistic electrons in the random and small-scale magnetic fields \citep{medvedev00,medvedev06,sirony09,fre10,kelner13}, is an appropriate radiation mechanism to study the stochastic polarization.
In our former works, we proposed that the turbulence is originated from the random and small-scale magnetic fields, and this assumption was realized by hydrodynamic simulations \citep{sch04}. We applied the jitter radiation to reproduce the GRB prompt emission \citep{mao11}.
We utilized the jitter self-Compton scattering and obtained the GRB emission in the GeV-band \citep{mao12}. In particular, we adopted the jitter radiation to reproduce the GRB prompt polarization in the high-energy band \citep{mao13}. The stochastic magnetic fields have random distribution, such that the relativistic electrons have same Lorentz force on average from different directions. Thus the jitter radiation is highly symmetric in the electron radiative plane. The net polarization degree should be zero. However, in our work, we proposed that the relativistic shocks compress a three-dimensional magnetic cube into a two-dimensional slab. Although the magnetic fields in the two-dimensional slab have random distribution, this slab provides a certain asymmetric configuration \citep{laing80,laing02}. Therefore we may detect the highly polarized jitter radiation.

We further study the depolarization properties and the circular polarization features using jitter radiation in this work. This is a consequent paper from our former studies. We illustrate our model in Figure 1. Several self-consistent physical processes are included in our model. Turbulence is produced behind GRB shock front. The turbulence is originated from the random and small-scale magnetic fields. Jitter radiation spectrum is dominated by the turbulent energy spectrum. A three-dimensional turbulent cube is compressed into a two-dimensional slab by the relativistic shocks. With this certain magnetic field topology \citep{laing80,laing02}, intrinsic Stokes parameters can be simplified. The Faraday RM can be structured by a turbulent cascade \citep{tribble91},
and we can further determine the depolarization properties. The jitter photons may also scatter with those turbulent magnetic clumps inside the slab, and this turbulent slab can be treated as a birefringence. Thus the stochastic circular polarization can also be produced \citep{macquart00a,macquart00b}. In the optical band, these linearly and circularly polarized photons can be strongly absorbed by the dense medium inside the slab. Finally, the realistic Stokes parameters are obtained by solving the polarized radiation transfer equation.

We present a brief review of jitter radiation and turbulent property in Section 2.1. In the case of the special magnetic field configuration, the jitter polarization properties are given in Section 2.2. The wavelength-dependent depolarization is shown in Section 2.3. The circular polarization produced by the turbulent slab is estimated in Section 2.4. The solutions of the polarized radiation transfer equation are written in Section 2.5. In Section 3, we select three GRB cases with the optical polarization measurements and compare those observations with our modeling results. We summarize our results and present some discussion issues in Section 4.

\section{Polarization Processes}
The jitter radiation and the stochastic magnetic fields for the explanation of GRB high-energy emissions have been studied in our former works \citep{mao11,mao12,mao13}. These issues are reviewed in Section 2.1. The wavelength-dependent depolarization can be measured by the Faraday turbulent RM. The estimation of the turbulent circular polarization variance can be linked to the GRB fireball Lorentz factor and the Lorenz factor of the turbulent eddy. The solution of the radiation transfer equation provides the realistic Stokes parameters, which can be compared with some observational results in the optical band.
During our calculations, we neglect the polarization contribution of the interstellar medium in our Galaxy \citep{serkowski75,martin90}.

\subsection{Jitter Radiation with Turbulent Feature}
The relativistic electron radiation in the small-scale magnetic fields was initially mentioned by \citet{landau71}. Here, we follow the procedure given by \citet{mao11}, and we write the jitter radiation intensity (energy per unit frequency per unit time) of a single relativistic electron as
\begin{equation}
I_\omega=\frac{e^4}{m^2c^3\gamma^2}\int^{\infty}_{1/2\gamma_\ast^2}d(\frac{\omega'}{\omega})(\frac{\omega}{\omega'})^2{(1-\frac{\omega}{\omega'\gamma_\ast^2}
+\frac{\omega^2}{2\omega'^2\gamma_\ast^4})}
\int{dq_0d{\bf q} \delta(w'-q_0+{\bf qv})K({\bf q})\delta[q_0-q_0({\bf q})]},
\end{equation}
where $\gamma_\ast^{-2}=(\gamma^{-2}+\omega^2_{pe}/\omega^2)$,
$\omega'=(\omega/2)(\gamma^{-2}+\theta^2+\omega^2_{pe}/\omega^2)$,
$\omega_{pe}=(4\pi e^2n/\Gamma_{sh}m_e)^{1/2}$ is the plasma frequency, $\Gamma_{sh}$ is the bulk Lorentz factor of the shock, $\theta\sim 1/\gamma$ is defined as the angle between the electron velocity and the radiation direction, and $\gamma$ is the electron Lorentz factor.
The dispersion relation of $q_0=q_0(\bf{q})$ is in the fluid field, q is the wavenumber of the disturbed fluid field, $q_0$ is the frequency of the disturbed fluid field, $v$ is the electron velocity from the perturbation theory, and we use the relation of $\omega'=q_0-\bf{qv}$ to link the radiation field to the fluid field.
The dispersion relation of the relativistic collisionless shocks was given by \citet{mi06}, and
we further obtain $q_0=cq[1+\sqrt{1+4\omega_{pe}^2/\gamma c^2q^2}/2]^{1/2}$.

The stochastic magnetic field $\langle B(q)\rangle$ generated by the turbulent cascade can be given by
$K(q)\sim \langle B^2(q)\rangle\sim \int_q ^{\infty} F(q')dq'$,
where $F(q)\propto q^{-\zeta_p}$. The number of $q$ is within the range of $q_\nu<q<q_\eta$.
The parameter $q_\nu$ is linked to the turbulent viscosity, and the parameter $q_\eta$ is related to the resistive transfer. Although the magnetic energy strongly increases in the length scale of $q_\nu<q<q_\eta$, it is reasonable that we use the length scale of $l_\nu$ as the typical scale for the magnetic field generation \citep{schekochihin07}.
Thus the magnetic field can be
$\langle B\rangle\sim q_{\nu}^{(1-\zeta_p)/2}/\sqrt{\zeta_p-1}$.
Through the turbulent cascade processes, the turbulent energy shows a hierarchical fluctuation structure, and
the inertial-range scaling laws of the fully developed turbulence were presented as
$\zeta_p=p/9+2[1-(2/3)^{p/3}]$ \citep{she94,she95}, where the energy spectrum index $\zeta_p$ of the
turbulent field is related to the cascade process number $p$. The Kolmogorov turbulent spectrum index is presented as $\zeta_p=p/3$.
Therefore we solve Equation (1) and obtain the radiation property $I_\omega\propto \omega^{-(\zeta_p-1)}$.
We emphasize that the spectral index $\zeta_p-1$ of this simplified jitter radiation is related to the fluid turbulence.

\subsection{Intrinsic Linear Polarization}
The polarization processes presented in this subsection have been fully studied by \citet{mao13}. Here we simply write some major results.
By the relativistic shocks, a three-dimensional cube is compressed into a two-dimensional slab.
The slab contains the small-scale magnetic fields, and the magnetic fields are stochastic \citep{laing80,laing02}.
The magnetic field vector at any point in the slab plane is described as
${\bf B}=B_0(\cos\phi \sin\theta_B, \sin\phi, \cos\phi \cos\theta_B)$, where
$\theta_B$ is defined as the angle between the slab plane and the line-of-sight (see Figure 1), and
$\phi$ is the azimuthal angle in the slab plane.

The position angle of the $E$-vector $\chi$ is $\cos 2\chi=-(\sin^2\theta_B-\tan^2\phi)/(\sin^2\theta_B+\tan^2\phi)$.
Thus the magnetic field for the radiation is $B=B_0(\cos^2\phi \sin^2\theta_B+\sin^2\phi)^{1/2}$.
The acceleration term related to the jitter radiation is proportional to $B^2$.
We take this magnetic field topology to calculate the decomposition of the jitter radiation on the electron radiation plane, and we finally obtain the polarization degree.
We orientate the slab to the case that the $E$-vector of the polarized radiation has the position angle of zero degrees. Thus we obtain the Stokes parameter $U=0$.
The left-side circular polarization can exactly cancel out the right-side circular polarization. Then, we have $V=0$ (see Appendix A of \citet{mao13} for the detailed calculation).
Thus, the Stokes parameters of the single electron jitter radiation are
$I=C\int_0^{2\pi}(\cos^2\phi \sin^2\theta_B+\sin^2\phi)d\phi$,
and
$Q=I\rm{cos}2\chi=-C\int_0^{2\pi}(\cos^2\phi \sin^2\theta_B-\sin^2\phi)d\phi$,
where $C=C(\gamma, B_0)$.
With a certain electron energy distribution $N_e(\gamma)$, we obtain the Stokes parameters from the jitter radiation of total electrons as
\begin{equation}
I_0=\int^{\gamma_2}_{\gamma_1}C(\gamma, B_0)N_e(\gamma)d\gamma \int_0^{2\pi}(\cos^2\phi \sin^2\theta_B+\sin^2\phi)d\phi,
\end{equation}
\begin{equation}
Q_0=-\int^{\gamma_2}_{\gamma_1}C(\gamma, B_0)N_e(\gamma)d\gamma \int_0^{2\pi}(\cos^2\phi \sin^2\theta_B-\sin^2\phi)d\phi,
\end{equation}
\begin{equation}
U_0=0,
\end{equation}
and
\begin{equation}
V_0=0.
\end{equation}
The linear polarization degree of the jitter emission is
\begin{equation}
\Pi_{l,0}=\frac{Q_0}{I_0}=-\frac{\int_0^{2\pi}(\cos^2\phi \sin^2\theta_B-\sin^2\phi)d\phi}{\int_0^{2\pi}(\cos^2\phi \sin^2\theta_B+\sin^2\phi)d\phi}.
\end{equation}
The magnetic field configuration is a key point to determine the jitter polarization in our model. Therefore the jitter polarization degree is tightly related to $\theta_B$. Here we do not consider the coherence of the turbulent magnetic field \citep{pro16}.

\subsection{Wavelength-dependent Depolarization}
The two-dimensional turbulent slab where the emission at wavelength $\lambda$ has the Faraday RM and makes an effective depolarization. This depolarization is wavelength-dependent. In this work, we adopt the results given by \citet{tribble91}. The detailed presentation is listed in Appendix A. Here we consider the slab, which has the turbulent feature with a structure of
\begin{equation}
D(s)=2\sigma^2(s/r_1)^{\zeta_p},
\end{equation}
where $r_1\sim 1/q_\nu$ is the outer scale of the turbulence, $\sigma$ is the dispersion of a Gaussian-distributed random field RM, and $\zeta_p$ is the index of a turbulent spectrum. This structure function represents the stochastic magnetic field due to the turbulent cascade. After the calculations presented in Appendix A, we obtain the polarization, which is related to the observational wavelength $\lambda$ as
\begin{equation}
\Pi(\lambda)=\Pi_0(\frac{\lambda}{\lambda_0})^{-4/\zeta_p}.
\end{equation}
Compared with the expression $\Pi(\lambda)\propto exp(-\lambda^4)$ given by \citet{burn66}, the depolarization presented in Equation (8) has a much slower fall-off. Here we do not make a simple tautology of \citet{tribble91}. We point out in this work that
the jitter depolarization feature is directly related to the turbulent energy spectrum and the jitter radiation spectrum. Because this depolarization feature takes effect on the long-wavelength observations, we expect the optical and radio depolarization measurements in the future.
Here we neglect the complicated depolarization effect from our Galaxy \citep{sokoloff98}.

\subsection{Turbulence-induced Circular Polarization}
The intrinsic circular polarization cannot be generated by the former magnetic field topology in Section 2.2. However, if the jitter photons are scattered with those magnetized clumps in the turbulent slab, the nonzero variance of the Stokes parameter $V$ can be produced. This slab can be treated as a birefringence. Here we consider those clumps as the coherent patches with a size of $r_{\rm{diff}}$, which are within a refractive slab with a size of $r_{\rm{ref}}$ \citep{macquart00b}. Then the stochastic variance of the Stokes parameter $V_0$ is presented as $\langle V_0^2\rangle\sim D(r_{\rm{ref}})\langle I_0\rangle^2$, where the structure function is $D(r_{\rm{ref}})=\alpha^2(r_{\rm{ref}}/r_{\rm{diff}})^{\zeta_p}$. In the synchrotron radiation case, we have $\alpha=\nu_B\cos\theta/\nu$, where $\nu_B$ is the electron cyclotron radius and $\theta$ is the angle between the magnetic field and the electron moving direction. In our model, we use the jitter radiation instead of the synchrotron radiation to produce the polarization features.
Therefore we take the plasma frequency $2\pi\omega_{pe}$ instead of $\nu_B\cos\theta$ to calculate $\alpha$, and we present the stochastic circular polarization as
\begin{equation}
\frac{\langle V_0^2\rangle}{\langle I_0\rangle^2}=\alpha^2(\frac{r_{\rm{ref}}}{r_{\rm{diff}}})^{\zeta_p}.
\end{equation}
Here we take the refractive length scale $r_{\rm{ref}}$ as the size of the turbulent slab, which can roughly correspond to the GRB fireball radius $R$. The diffractive length scale $r_{\rm{diff}}$
corresponds to the scattering scale, which is less than the thickness of the fireball $\Delta R$. We estimate $r_{\rm{diff}}=\Delta R/\Gamma_t$, where $\Gamma_t$ is the bulk Lorentz factor of the turbulence.
With the relation $\Delta R/R=1/\Gamma_{sh}^2$ of the relativistic fireball model \citep{blandford76,piran99}, we obtain
\begin{equation}
\sqrt{\frac{\langle V_0^2\rangle}{\langle I_0\rangle^2}}=\frac{2\pi\omega_{pe}}{\nu_{obs}}\Gamma_t^{\zeta_p/2}\Gamma_{sh}^{\zeta_p},
\end{equation}
where $\omega_{pe}=(4\pi e^2n/\Gamma_{sh}m_e)^{1/2}$.
In the soft X-ray band, we obtain the stochastic circular polarization as
\begin{equation}
\sqrt{\frac{\langle V_0^2\rangle}{\langle I_0\rangle^2}}=2.6\times 10^{-6}(\frac{n}{3.0\times 10^{10}~\rm{cm^{-3}}})^{1/2}(\frac{\nu_{obs}(1+z)}{1.0~\rm{keV}})^{-1}\Gamma_t^{\zeta_p/2}\Gamma_{sh}^{(\zeta_p-1/2)}.
\end{equation}
If we take $\Gamma_t=10$, $\Gamma_{sh}=50$ (the internal shock case treated as the tail of the GRB prompt emission), $\zeta_p=2.0$, and $z=0$, we obtain a number of about 1.0\%. Although it is necessary to use the sensitive polarimeters to detect these weak signals, it seems that the soft X-ray GRB circular polarization is not negligible. We can also obtain the stochastic circular polarization in the optical band as
\begin{equation}
\sqrt{\frac{\langle V_0^2\rangle}{\langle I_0\rangle^2}}=2.0\times 10^{-4}(\frac{n}{3.0\times 10^{10}~\rm{cm^{-3}}})^{1/2}(\frac{\lambda_{obs,R}/(1+z)}{6400~\AA})\Gamma_t^{\zeta_p/2}\Gamma_{sh}^{(\zeta_p-1/2)}.
\end{equation}
If we take $\Gamma_t=10$, $\Gamma_{sh}=10$ (the external/reverse shock case), $\zeta_p=2.0$ and $z=0$, we obtain a number of about 6.3\%, which can be effectively detected by the optical polarimeters.

Because the stochastic circular polarization is induced by the turbulence, the high-degree circular polarization is originated from the
large number of $\Gamma_t$. We see that a steep GRB spectrum produced by a steep turbulent energy spectrum can make strong stochastic circular polarization. We adopt the physical concept of the turbulent screen given by \citet{macquart00b}, and we note the dense medium compressed by the relativistic shocks in the slab have a larger number density than the normal interstellar medium. If a large number density and a strong turbulence with a steep power-law energy spectrum are given, the stochastic circular polarization may be effectively detected. Thus the polarization degree predicted in our model can be much larger than that obtained from the interstellar medium scintillation in GRB host galaxies.
In our model,
the stochastic circular polarization
is not related to the view angle $\theta_B$,
and we caution that the circular polarization degree is strongly dependent on the fireball dynamical factor $\Gamma_{sh}$.

\subsection{Radiation Transfer Processes}
The GRB optical photons can be strongly absorbed by the dense medium in the slab. In order to provide the realistic polarization degrees of the GRB afterglow and compare with the observational data, it is necessary to calculate the absorbed Stokes parameters. In principle, we should solve the radiation transfer equation, and two aspects are included. One aspect is on the absorption term, and the other is on the Faraday rotation/conversion terms. In our model, the intrinsic linear polarization can be produced in a certain magnetic field topology, and the circular polarization is originated from the scattering with the magnetic clumps in the turbulent slab.
Because the stochastic Faraday RM effect is accounted for the circular polarization, we only consider the absorption term in the radiation transfer equation, and we neglect the rotativity and convertibility (``crosstalk") terms during the radiation transfer process.

One complete presentation of the radiation transfer processes is given in Appendix B. Here we write three equations for the Stokes parameters, in which only the absorption effect is considered:
\begin{equation}
\frac{dI}{d\tau}=-I-\frac{k_Q}{k_I}Q-\frac{k_V}{k_I}V,
\end{equation}
\begin{equation}
\frac{dQ}{d\tau}=-\frac{k_Q}{k_I}I-Q,
\end{equation}
and
\begin{equation}
\frac{dV}{d\tau}=-\frac{k_V}{k_I}I-V,
\end{equation}
where $d\tau=k_Idl$ is the optical depth of the absorption, and $k_I$, $k_Q$, and $k_V$ are the absorption coefficients given in Appendix B. We can solve these equations and obtain
\begin{equation}
I=C_2exp[-(1-\sqrt{(k^2_Q+k^2_V)/k^2_I})\tau]+C_3exp[-(1+\sqrt{(k^2_Q+k^2_V)/k^2_I})\tau],
\end{equation}
\begin{eqnarray}\nonumber
Q=C_1exp(-\tau)-C_2\frac{k_Q}{\sqrt{k_Q^2+k^2_V}}exp[-(1-\sqrt{(k^2_Q+k^2_V)/k^2_I})\tau]\\
+C_3\frac{k_Q}{\sqrt{k_Q^2+k^2_V}}exp[-(1+\sqrt{(k^2_Q+k^2_V)/k^2_I})\tau],
\end{eqnarray}
and
\begin{eqnarray}\nonumber
V=-C_1\frac{k_Q}{k_V}exp(-\tau)-C_2\frac{k_V}{\sqrt{k_Q^2+k^2_V}}exp[-(1-\sqrt{(k^2_Q+k^2_V)/k^2_I})\tau]\\
+C_3\frac{k_V}{\sqrt{k_Q^2+k^2_V}}exp[-(1+\sqrt{(k^2_Q+k^2_V/k^2_I})\tau],
\end{eqnarray}
where
\begin{equation}
C_2=\frac{1}{2}I_0-\frac{k_QQ_0+k_VV_0}{2\sqrt{k^2_Q+k^2_V}},~C_1=\frac{k_Q}{\sqrt{k^2_Q+k^2_V}}(2C_2-I_0)+Q_0,~\rm{and}~C_3=I_0-C_2.
\end{equation}
Thus the linear polarization $\Pi_l=Q/I$ and the circular polarization $\Pi_c=V/I$ can be determined. The polarization angle is also dependent on the absorption coefficients, as described in Appendix B.

We can further calculate a ratio between the circular polarization and the linear polarization. This ratio, derived from Equations (16)-(18), is not related to the optical depth. We plot the ratio as a function of the view angle $\theta_B$ in Figure 2. If the view angle $\theta_B$ is less than about 70$^\circ$, the circular polarization is weaker than the linear polarization. However, the circular polarization can be stronger than the linear polarization if the view angle $\theta_B$ is larger than 70$^\circ$. In the latter case, from the observational point of view, both the circular polarization and the linear polarization may be very weak. Thus some precise examinations from a sensitive optical polarimeter are necessary to constrain our results.

\section{Case Studies: GRB 091018, GRB 121024A, and GRB 131030A}
We qualitatively examine three cases of the optical/infrared GRB polarization in this Section.
Many high-quality data
of GRB 091018 were given, and the detections of both the linear polarization and the circular polarization were performed \citep{wier12}. The linear polarization and circular polarization results of GRB 121024A were also found by \citet{wier14}.
The linear polarization of GRB 131030A was considered as an origin of some disordered magnetic fields with the reverse shock \citep{king14}.

The observational polarization properties of GRB 091018 were presented in detail by \citet{wier12}. The linear polarization degree in the $R$-band is about 1.4\%, and the combined linear polarization degree in the $K_s$-band is 2\%.
The spectral index in the optical band is $\beta=0.58$, providing $\zeta_p=\beta+1=1.58$ in our model.
The depolarization feature derived in our model provides 
$\Pi_l(K_s)/\Pi_l(R)=(\lambda_{K_s}/\lambda_R)^{-4/1.58}=0.04$, much smaller than the observational one\footnote{We cannot exclude some more complicated cases: twisted magnetic fields make a polarization increasing with wavelength \citep{sokoloff98}, and/or depolarization can be originated from magnetohydrodynamic turbulence \citep{eilek89}.}. However, we note that the observational polarization degree in the $K_s$-band is a combined number, as the source in the $K_s$-band is fairly faint. This prevents us from further investigations.
On the other hand, in our model, the circular polarization is strongly dependent on the ionized plasma density $n$, which is quite uncertain. Here we assume that the medium can be fully ionized by the shock collision/propagation and its density can be viewed as a lower-limit of the plasma density affected on the circular polarization. The column density of the medium is $N_{\rm{H}}=1.8\times 10^{21}A_V~\rm{cm^{-2}}$, and $R_V=A_V/E(B-V)$. Here, $A_V$ and $E(B-V)$ are the intrinsic absorption from the materials in the slab and $R_v=3.1$.
For the case of GRB 091018, the intrinsic extinction from the observation is $E(B-V)=0.024$, and the hydrogen column density is $N_{\rm{H}}=1.3\times 10^{20}~\rm{cm^{-2}}$.
If we assume the plasma density $n\sim n_{\rm{H}}$, and we take the thickness of a fireball shell as the absorption length scale of about $1.0\times 10^9$ cm, we roughly estimate $n\sim 1.3\times 10^{11}~\rm{cm^{-3}}$. Finally, we predict the intrinsically stochastic circular polarization degree from Equation (12) as 1.6\% when we adopt $\Gamma_{sh}=10.0$ and $\Gamma_t=10.0$. However, the observation only provides an upper limit of 0.15\%. Therefore, for this GRB, it is difficult to have any reliable conclusion from our model.

The detections of the linear polarization (4\% degree) and the circular polarization (0.6\% degree) in GRB 121024A have been carried out in the optical band \citep{wier14}. The X-ray to optical power-law spectrum provide a spectral index of $\beta=0.88$, corresponding to the index of turbulent energy spectrum of $\zeta_p=\beta+1=1.88$ in our model.
The intrinsic absorption from the X-ray to optical spectral energy distribution is $A_V=0.22$, corresponding to a hydrogen column density of $N_{\rm{H}}=4.0\times 10^{20}~\rm{cm^{-2}}$. If we take the absorption length scale of about $1.0\times 10^9$ cm, we obtain $n\sim n_{\rm{H}}\sim 4.0\times 10^{11}~\rm{cm^{-3}}$. We apply Equation (12) with $\Gamma_{sh}=10.0$ and $\Gamma_{t}=10.0$, and the intrinsically stochastic circular polarization degree can be 4.6\%. From our model, the ratio between the circular polarization and the linear polarization is not related to the optical depth of the absorption. Thus, from the lower-right panel in Figure 2, we can constrain the view angle as $\theta_B\sim 40^\circ$. If we consider both the waveband depolarization and the strong absorption of the dense medium as $\tau\sim 0.5$, we may roughly reproduce the observational numbers of the linear polarization degree and the circular polarization degree. It seems that it will be possible to explain the observational polarization features of GRB 121024A with our model.

The low-degree (2.1\%) linear polarization of GRB 131030A has been suggested as an origin of some disordered magnetic fields from the plasma instability in the ambient interstellar medium \citep{king14}. The magnetic fields and the related polarization induced by the  Weibel instability were given by \cite{med99}. If the disordered magnetic fields are considered, the jitter radiation should be applied. In our model, the special magnetic field topology provides the possibility to intrinsically reproduce either the high-degree or the low-degree linear polarization, while another possibility is that the intrinsic high-degree linear polarization is strongly depolarized and heavily absorbed by the dense medium. Here we do not have the X-ray to optical spectrum. The photon index and the intrinsic column density (at $z=1.293$) from the X-ray spectrum\footnote{http://www.swift.ac.uk/xr$\rm{t}_-$spectra/00576238/} are $\Gamma_x=1.93$ and $N_{\rm{H,X}}=4.2\times 10^{21}~\rm{cm^{-2}}$, respectively. It seems that the depolarization and the absorption cannot be easily neglected. However, we have no multi-wavelength spectrum to further constrain our model, and we cannot obtain any reliable conclusion for GRB 131030A.

\section{Discussion and Conclusions}
Turbulence is a key point in our model. A three-dimensional magnetic cube can be compressed as a two-dimensional slab by the relativistic shocks. This topology of the random magnetic fields provides the jitter radiation properties. The turbulence not only induces the random and small-scale magnetic fields, but also provides a structured RM in this plasma screen. Thus the wavelength-dependent depolarization and the stochastic circular polarization are produced. We also speculate that the time-variable linear polarization observed in the high-energy and optical bands \citep{greiner03,gotz09,yonetoku11,wier12,mundell13}
may be originated from the turbulence.
Several instances of progress have been noted in recent years. The turbulent microphysics can be estimated (see \citet{lemoine13} and references therein), and the relativistic turbulence has the same spectral and intermittency properties of the non-relativistic turbulence \citep{zrake13}. The radiation with Langmuir turbulence can be numerically examined \citep{teraki13}.
Some observational pulses with the short-timescale variances shown in the {\it Swift}-BAT GRB lightcurves indicate the turbulence feature \citep{bhatt12}.
Thus the stochastic magnetic field \citep{laing80,laing02} and the turbulent birefringence \citep{macquart00a,macquart00b} can be naturally employed in our model. All of these physics points, such as shocks, turbulence, jitter radiation, depolarization, and radiation transfer, are self-consistent in our framework. We note that the physical scenario in this paper could be also applied in the polarization study of blazars \citep{mead90,ikejiri11,cha15,covino15} and fast radio bursts (FRBs; Lorimer et al. 2007; Petroff et al. 2015; Chatterjee et al. 2017). For example, if the conversion terms in the radiation transfer equation are considered in the three-dimensional blazar jet, we expect an extra circular polarization as the jet propagates. The emission terms in the radiation transfer equation are also important for the blazar jet. These possibilities are realized recently by some numerical tests (MacDonald \& Marscher 2017). The three-dimensional model of the time-dependent jitter polarization and radiation transfer can be improved in the future.

One may infer that the low-degree linear polarization in some optical observations can reveal some features of the disordered magnetic fields \citep{gruzinov99,king14,wier14}. In our model, the jitter radiation can produce either the high-degree or the low-degree linear polarizations in the optical band. The stochastic circular polarization of at least a few percent can be produced by the jitter radiation as well if the shock-swept turbulent plasmas are dense.
Moreover, reverse shock \citep{meszaros99,sari99,kobayashi00,fan02,zhang03} has been proposed for the early brightening of a GRB afterglow, and it may take effects on the optical and radio polarizations \citep{granot05,king14}. Here we suggest that the reverse shock may make strong ionization, naturally produce strong turbulence, and effectively pile up materials. Therefore the reverse shock can produce the strong circular polarization accompanied by the linear polarization in the optical, infrared, and radio bands, although these polarized photons may suffer a strong absorption.

The relativistic electron radiation in the stochastic magnetic fields is treated by the jitter mechanism in our model. With the certain magnetic field topology, either the high-degree or the low-degree intrinsic linear polarization can be reproduced. Thus we can explain the high-degree linear polarization in the high-energy band.
Some ongoing and future high-energy polarimeters, such as POLAR \citep{orsi12}, ASTROSAT \citep{paul13}, TSUBAME \citep{yatsu14}, HARPO \citep{wang15}, and polSTAR \citep{kra16}, can effectively identify some polarized GRBs.
We also expect that the optical and radio polarization detections at different wavelengths can directly examine the GRB depolarization effect.
GRBs alerted with the initial optical counterparts by {\it Swift} are considered as good candidates for the optical polarization detections (e.g., GRB 060418, GRB 090102, GRB 091018, GRB 091208B, GRB 110205A, GRB 120308A, GRB 131030A, etc.), because optically bright sources are efficient to perform the polarization measurements from the observational point of view.
In particular, the radio observations are relatively rare for the examination of the GRB polarization properties. An upper limit of the linear polarization of about 2\% in the radio band was set for GRB 030329 \citep{taylor05}. The parameters of the circular polarization presented in our model, such as the number density of plasma, Lorentz factors of shock and turbulence, and absorption coefficient, can be further constrained by the radio observations. The circular polarization measurements of GRB 030329 in the radio band were performed by \citet{fink04}. However, the measurement errors of the circular polarizations were very large.
Thus more radio polarization measurements are strongly suggested. Moreover, we also suggest simultaneously multi-wavelength polarization measurements in the future to enhance the study of GRB physics. We hope that our results and expectations in this paper can be helpful for more kinds of polarization observations.

\acknowledgments
We are grateful to the referee for his/her careful review and suggestions.
J. Mao is supported by the Hundred Talent Program, the Major Program of the Chinese Academy of Sciences (KJZD-EW-M06), the National Natural Science Foundation of China 11673062, and the Oversea Talent Program of Yunnan Province.
J. Wang is supported by the Strategic Priority Research Program ``The Emergence of Cosmological Structures" of the Chinese Academy of Sciences (XDB09000000), and the National Natural Science Foundation of China (11573060 and 11661161010).

\appendix
\section{Wavelength-dependent Depolarization}
We apply the calculation processes of \citet{tribble91} to derive the depolarization properties if the polarization RM is dominated by the turbulent cascade. Although the original calculation was for radio observations, the derived RM structure function and the long-wavelength approximation are valid in our purpose. The polarization distribution with the telescope beam is
\begin{equation}
\Pi({\bf x_0},\lambda)=\int W({\bf x}_0-{\bf x})\epsilon({\bf x})exp(2iRM({\bf x})\lambda^2+2i\theta({\bf x}))d^2{\bf x},
\end{equation}
where $W({\bf x})$ is the telescope beam function, $\epsilon({\bf x})$ is the polarized intensity, and $\theta({\bf x})$ is the initial position angle. We obtain
\begin{equation}
\langle |\Pi(\lambda)|^2\rangle=\int R({\bf s})\xi({\bf s})d^2{\bf s},
\end{equation}
where
\begin{equation}
R({\bf s})=\int W({\bf x})W({\bf x}+{\bf s})d^2{\bf x},
\end{equation}
and
\begin{equation}
\xi({\bf s})=\xi_f({\bf s},\lambda)\xi_\epsilon({\bf s})\xi_\theta({\bf s}),
\end{equation}
if the emissivity and the angle are not correlated. We have $\xi_\epsilon({\bf s})=\langle\epsilon({\bf x})\epsilon({\bf x}+{\bf s})\rangle$ and
$\xi_\theta({\bf s},\lambda)=exp(-2\langle[\theta({\bf x})-\theta({\bf x}+{\bf s})]^2\rangle)$. Here we do not resolve filamentary structures on the small scale, and the intrinsic polarization angle is assumed to be constant. Thus we simply neglect the effect of $\xi_\epsilon$ and $\xi_\theta$. Then, we have only the effect of $\xi_f({\bf s})$, which can be presented as
\begin{equation}
\xi_f({\bf s},\lambda)=\langle exp[2iRM({\bf x})\lambda^2-2iRM({\bf x}+{\bf s})\lambda^2]\rangle=exp[-2\lambda^4D({\bf s})],
\end{equation}
where $D({\bf s})=\langle[RM({\bf x}+{\bf s})-RM({\bf x})]^2\rangle$ is the RM structure function. In our case, we have the turbulence in the small scales so that we present
\begin{equation}
D(s)=2\sigma^2[(s^2+r_0^2)/r_1^2]^{\zeta_p/2}\approx 2\sigma^2(s/r_1)^{\zeta_p}
\end{equation}
in the case of $s\gg r_0$, where $\sigma$ is the dispersion of Gaussian random field in the RM, and $r_1$ is the outer scale of the turbulence.
If we assume the telescope beam has a Gaussian function as
\begin{equation}
W({\bf x})=\frac{1}{\pi t^2}exp(-|{\bf x}|^2/t^2),
\end{equation}
where $t$ has the relation of $w=2t\sqrt{\rm{ln} 2}$, we obtain
\begin{equation}
R({\bf s})=\frac{1}{\pi^2t^4}\int exp(-|{\bf x}|^2/t^2)exp(-|{\bf x}+{\bf s}|^2/t^2)d^2{\bf x}=\frac{1}{2\pi t}exp(-|{\bf s}|^2/2t^2).
\end{equation}
Therefore, we derive
\begin{equation}
\langle |\Pi(\lambda)|^2\rangle=\int R({\bf s})exp(-2\lambda^4D({\bf s}))d^2{\bf s}\approx R(0)\int exp[-2\lambda^4\sigma^2(s/r_1)^{\zeta_p}]2\pi sds.
\end{equation}
Finally we obtain
\begin{equation}
\langle |\Pi(\lambda)|^2\rangle=\frac{r_1^2}{t^2}\frac{\Gamma(2/\zeta_p)}{\zeta_p}(4\sigma^2\lambda^4)^{-2/\zeta_p}.
\end{equation}
We clearly see that $\Pi(\lambda)\propto \lambda^{-4/\zeta_p}$, and $\zeta_p$ is the index of a turbulent energy spectrum.

\section{Absorption coefficients of Jitter Radiation and Radiation Transfer Equation}
The absorption coefficient in general can be calculated as \citep{rybicki79,workman08}
\begin{equation}
k_\nu=-\frac{c^2}{8\pi\nu^2}\int I_\omega(\nu,E) E^2\frac{\partial}{\partial E}[\frac{N_e(E)}{E^2}]dE=-\frac{1}{8\pi m_0\nu^2}\int I_\omega(\nu, \gamma) \gamma^2\frac{\partial}{\partial \gamma}[\frac{N_e(\gamma)}{\gamma^2}]d\gamma,
\end{equation}
where $N_e$ is the electron energy distribution. We can obtain the jitter intensity $I_\omega$ from Equation (1)
in Section 2.1.
Electrons are accelerated to have a power-law energy spectrum by the diffusive shock acceleration. Turbulence can generate magnetic fields and accelerate electrons. Then the electron energy distribution shows a Maxwellian shape \citep{stawarz08}.
Thus the final electron energy distribution presents a shape of \citep{stawarz08,giannios09}
\begin{equation}
 N_e(\gamma)= \left\{
\begin{array}{l l}
C_e\gamma^2\rm{exp}(\gamma/\Theta)/2\Theta^3, & \gamma \le \gamma_{th}, \\
C_e\gamma_{th}^2\rm{exp}(\gamma_{th}/\Theta)(\gamma/\gamma_{th})^{-p_e}/2\Theta^3, & \gamma> \gamma_{th}, \\
\end{array} \right.
\end{equation}
where $C_e$ is a constant, and $p_e$ is the power-law index of the electron energy distribution. Because a non-thermal distribution is for a fraction of electrons, we take characteristic temperature
$\Theta=kT/m_ec^2\sim 200$. In our calculation, we adopt $\gamma_{min}=100$, $\gamma_{th}=10^3$, and $p_e=2.2$.
For example, if we take the $R$-band observation where the observational wavelength is $\lambda=6400\AA$ and a single electron radiation intensity
$I_\omega=3.1\times 10^{-11}~\rm{erg~Hz^{-1}~s^{-1}}$, with the electron density of $C_e= 10.0~\rm{cm^{-3}}$ and the absorption length of $l_s\sim \Delta R\sim 1.0\times 10^9$ cm,
the absorption optical depth is $\tau= 0.22$. However, we simply take the optical depth $\tau$ as a free parameter in the radiation transfer calculation, and
$\tau< 1$ is for the optical-thin case.

We have the Stokes parameters of the single electron jitter radiation presented in Section 2.2 as: $I=C\int_0^{2\pi}(\cos^2\phi \sin^2\theta_B+\sin^2\phi)d\phi$, $Q=I\rm{cos}2\chi=-C\int_0^{2\pi}(\cos^2\phi \sin^2\theta_B-\sin^2\phi)d\phi$, and $U=0$, where $C=C(\gamma, B_0)$. We can also obtain $\sqrt{\langle V^2\rangle}$ from Section 2.4. Thus we obtain the absorption coefficients of the Stokes parameters from Equation (B1) as
$k_I=A(\gamma,B_0)\int_0^{2\pi}(\cos^2\phi\sin^2\theta_B+\sin^2\phi)d\phi/8\pi m_0\nu^2$,
$k_Q=A(\gamma,B_0)\int_0^{2\pi}(\cos^2\phi \sin^2\theta_B-\sin^2\phi)d\phi/8\pi m_0\nu^2$, $k_U=0$, and
$k_V=A(\gamma, B_0)(r_{\rm{ref}}/r_{\rm{diff}})^{\zeta_p/2}\alpha\int_0^{2\pi}(\cos^2\phi \sin^2\theta_B+\sin^2\phi)d\phi/8\pi m_0\nu^2$.
Finally, the ratios of $k_Q/k_I$, $k_V/k_I$, and $k_V/k_Q$ are not related to $A(\gamma, B_0)$.

We present the radiation transfer equation of the polarization as \citep{sazonov69a,sazonov69b,jones77,matsumiya03,sagiv04,toma08}
\begin{eqnarray}
\left(
\begin{array}{cccc}
d/dl+k_I & k_Q      & 0        & k_V \\
k_Q      & d/dl+k_I & k^\ast_V & 0 \\
0        &-k^\ast_V & d/ds+k_I & k^\ast_Q \\
k_V      & 0        & -k^\ast_Q& d/ds+k_I
\end{array}     \right)
\left(
\begin{array}{c}
I  \\
Q  \\
U  \\
V
\end{array}     \right)
= \left(
\begin{array}{c}
\eta_I  \\
\eta_Q  \\
\eta_U  \\
\eta_V
\end{array}     \right).
\end{eqnarray}
The polarization angle defined by $\rm{tan} 2\chi=(U/Q)$ is also dependent on the absorption coefficients, and it can be calculated in general.
In our model, we do not have any additional emissions from the dense medium in the turbulent slab. Thus the emission coefficients of $\eta_I$, $\eta_Q$, $\eta_U$, and $\eta_V$ can be zero. Here we only take the absorption effect. Thus the rotativity $k^\ast_V=0$ and convertibility $k^\ast_Q=0$. We also take $U=0$. Finally, the Equation (B3) can be written as
\begin{eqnarray}
\left(
\begin{array}{cccc}
d/dl+k_I & k_Q      & 0        & k_V \\
k_Q      & d/dl+k_I & 0 & 0 \\
0        & 0 & d/ds+k_I & 0 \\
k_V      & 0        & 0 & d/ds+k_I
\end{array}     \right)
\left(
\begin{array}{c}
I  \\
Q  \\
0  \\
V
\end{array}     \right)
= \left(
\begin{array}{c}
0  \\
0  \\
0  \\
0
\end{array}     \right).
\end{eqnarray}

\clearpage




\begin{figure}
\includegraphics[angle=0,scale=.4]{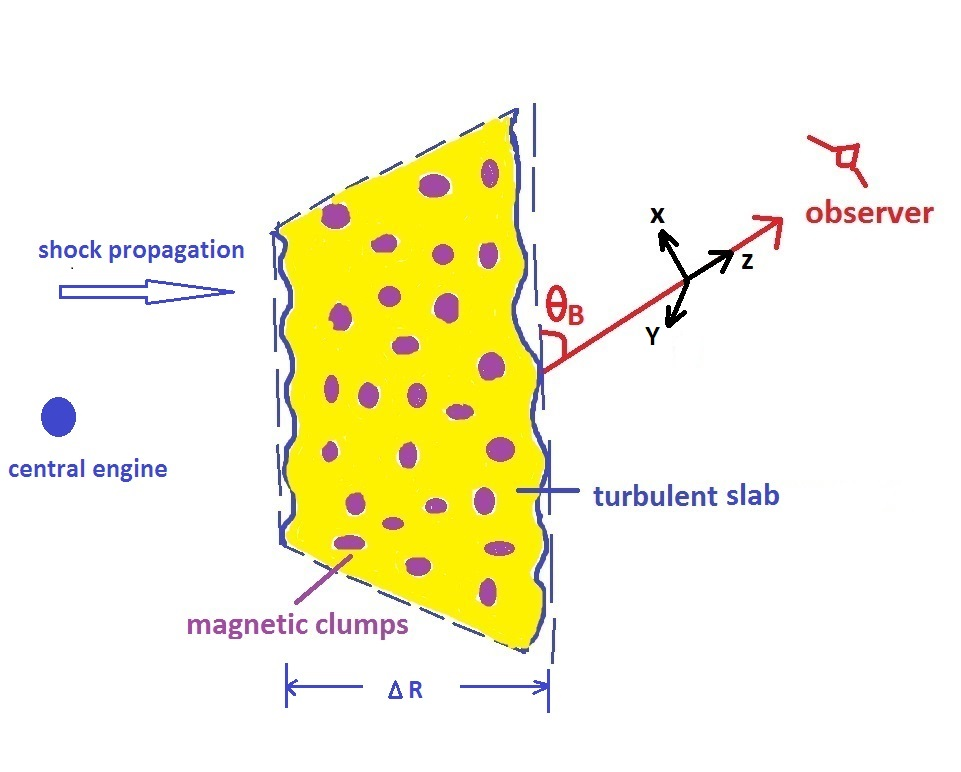}
\includegraphics[angle=0,scale=.4]{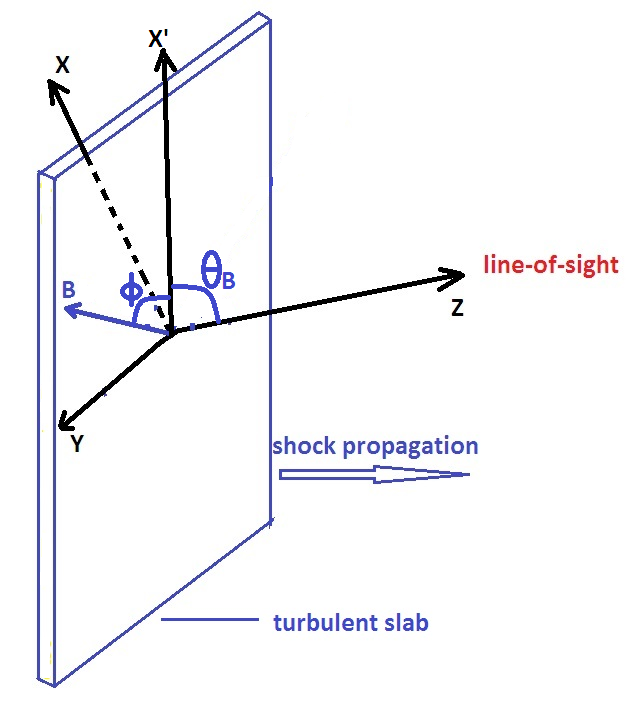}
\caption{Illustration of GRB polarization in our model. Left panel: a carton of a zoom-in turbulent slab. A three-dimensional cube near GRB central engine is compressed to a two-dimensional slab. The thickness of this slab $\Delta R$ is much smaller than the radius of the fireball $R$. The random and small-scale magnetic fields are generated by the turbulence in this slab. Linearly polarized jitter photons are produced in the turbulent magnetic clumps. In the meanwhile, the jitter photons scatter with those clumps, which have the diffractive scale $r_{\rm diff}$. The turbulent slab can be treated as a birefringence with a size of $r_{\rm def}\sim R$. Thus the stochastic circular polarization can be produced. The dense medium inside this slab can further absorb the polarized photons.
Right panel: an observer can detect the polarized GRB photons through the line-of-sight ($Z$-axis), which has the view angle of $\theta_B$ to the slab plane ($\bf{\rm{X'}}$-$Y$ plane). The magnetic field {\bf B} is inside the slab plane. Thus we have $\rm{B_{X'}=B_0cos\phi}$ and $\rm{B_Y=B_0sin\phi}$, where $\phi$ is the azimuthal angle at any point randomly distributed in the slab plane. $\rm{B_{\perp}}=(B_X,B_Y)$, which takes effect on the jitter radiation, is in the $X$-$Y$ plane, and we obtain $\rm{B_X=B_{X'}sin\theta_B}$. Finally we obtain $\rm{B_X=B_0\rm{cos}\phi\rm{sin}\theta_B}$ and $\rm{B_Y=B_0\rm{sin}\phi}$.
\label{fig1}}
\end{figure}

\clearpage

\begin{figure}
\includegraphics[scale=0.3]{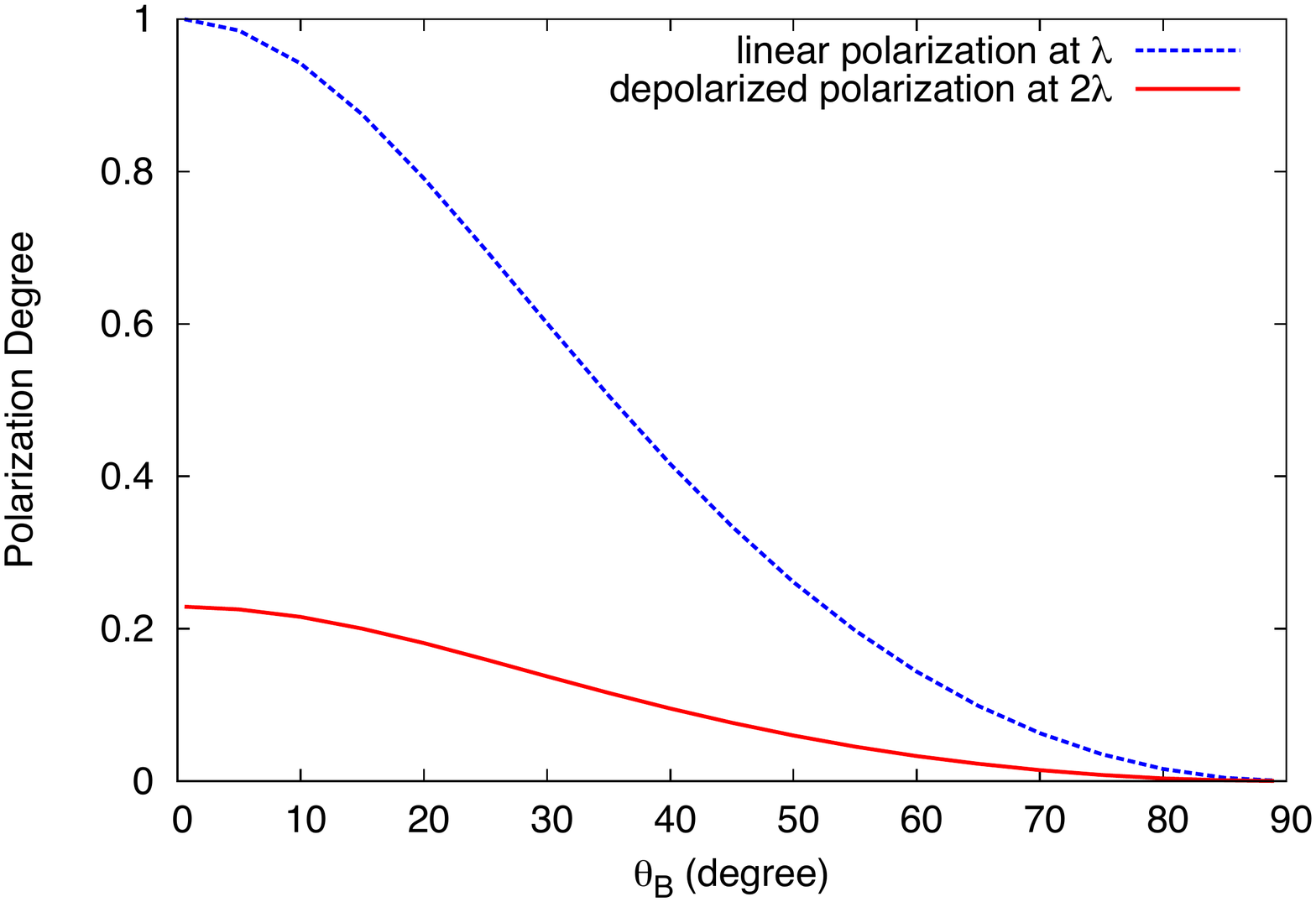}
\includegraphics[scale=0.3]{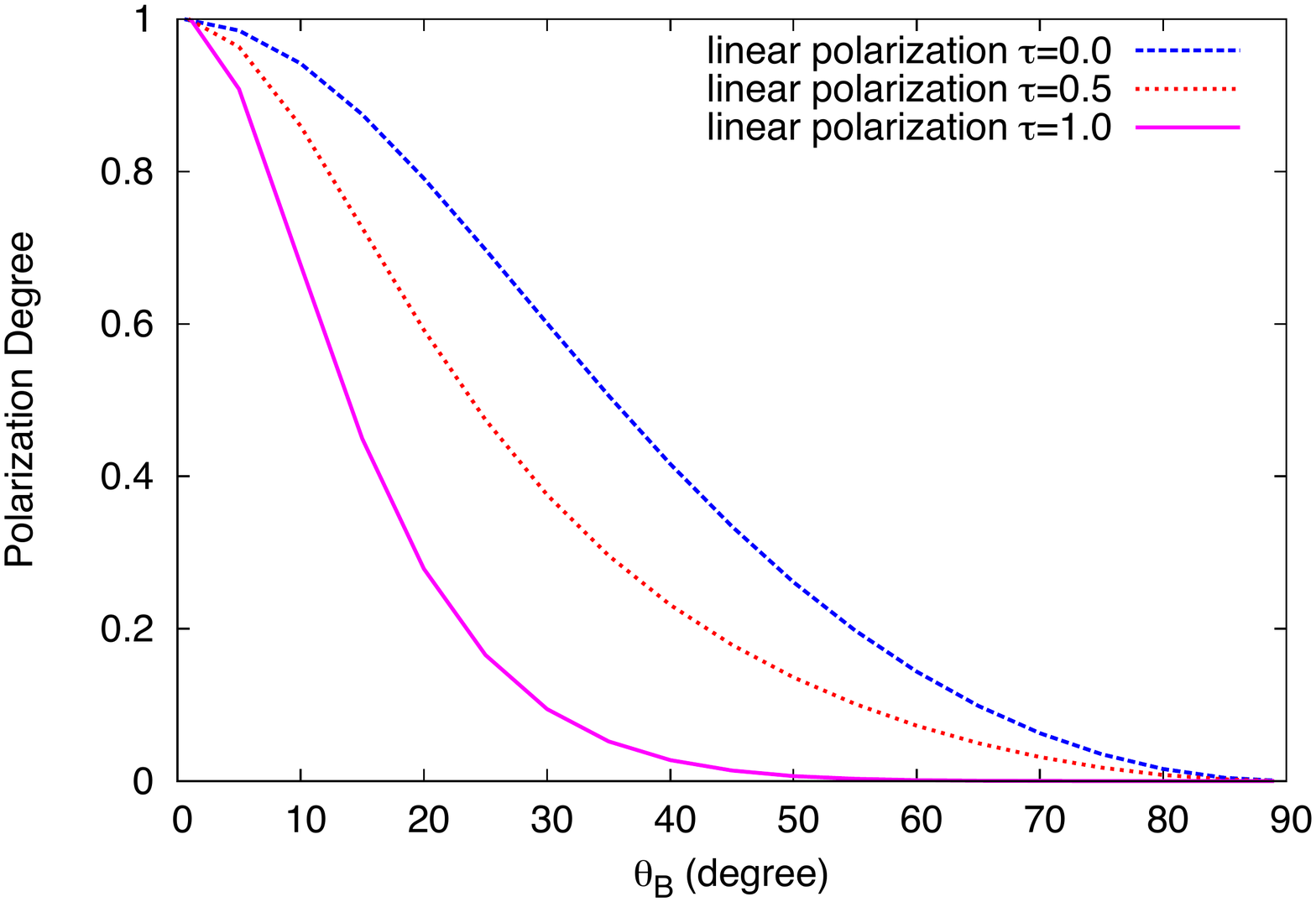}
\includegraphics[scale=0.3]{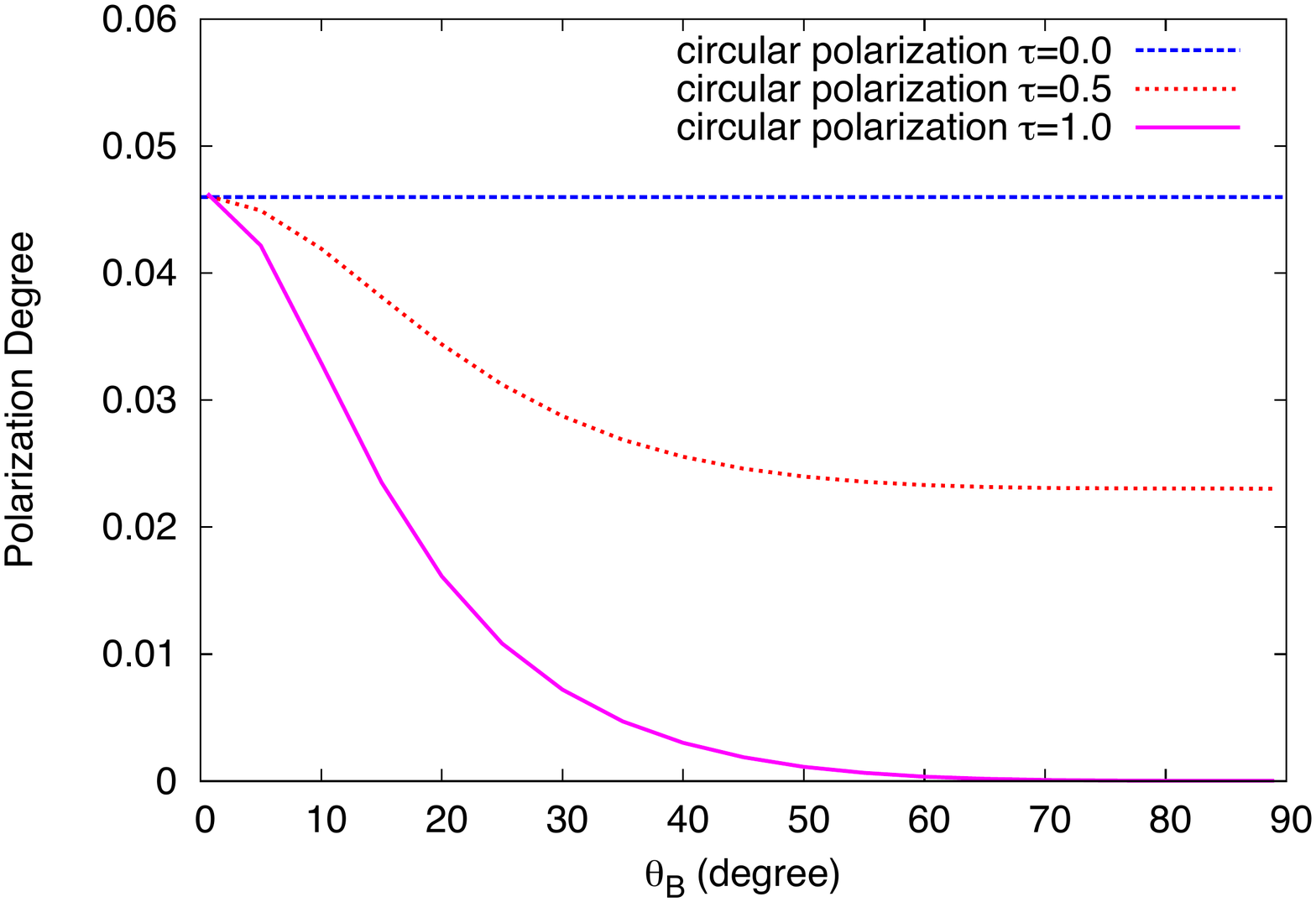}
\includegraphics[scale=0.3]{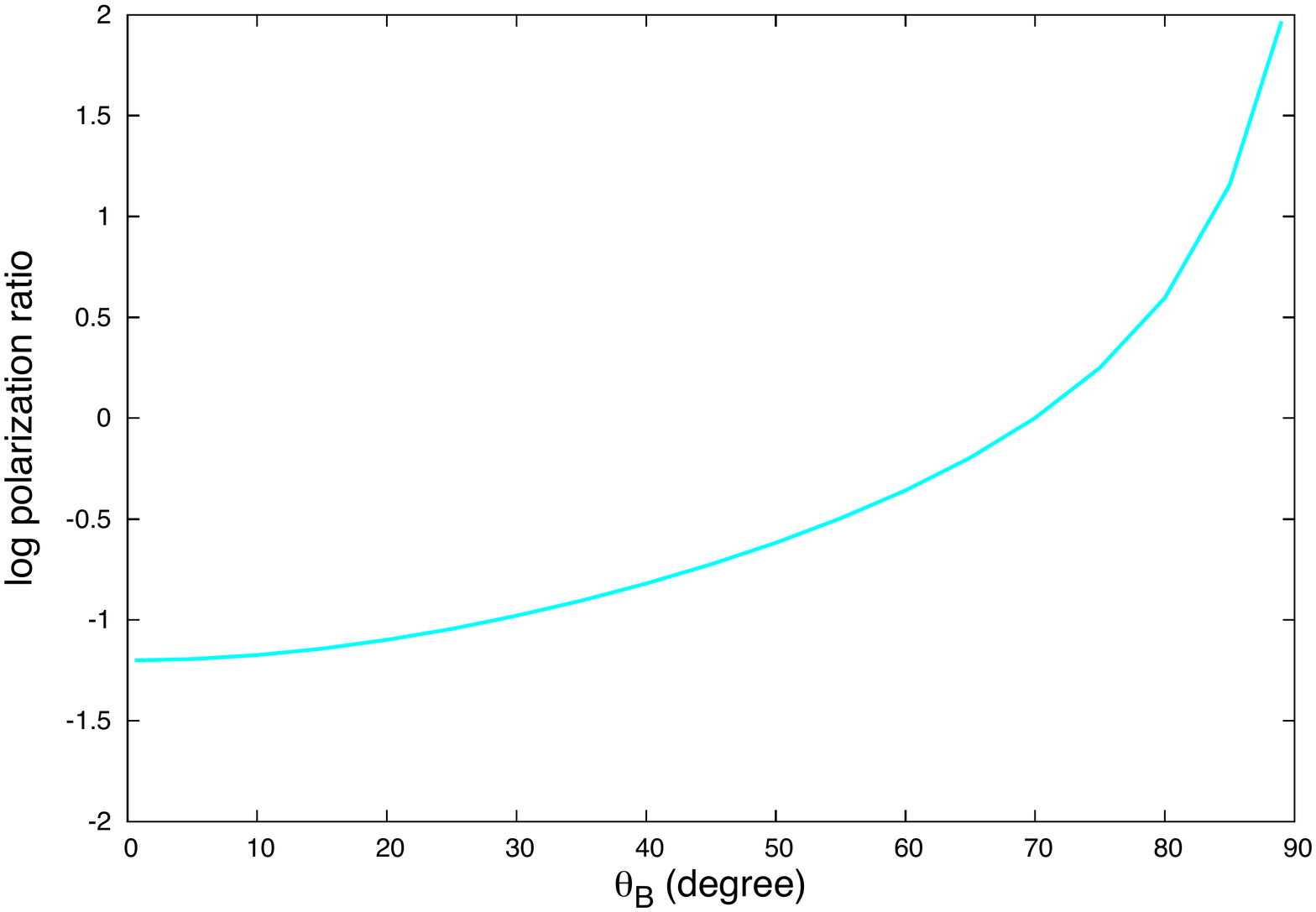}
\caption{Polarization degree as a function of $\theta_B$. Upper-left panel: the wavelength-dependent depolarization $\Pi\propto \lambda^{-4/\zeta_p}$. Here we take $\zeta_p=\beta+1=1.88$, as the spectral index $\beta=0.88$ of GRB 121024A, determined by \citet{wier14}. Upper-right panel: the linear polarization with the different absorption optical depth. Lower-left panel: the circular polarization with the different absorption optical depth. The results are obtained by the following parameters: $\Gamma_{sh}=10$, $\Gamma_t=10$, and $n=4.0\times 10^{11}~\rm{cm^{-3}}$. The spectral index $\beta=0.88$ of GRB 121024A \citep{wier14} is also adopted in the calculation. Lower-right panel: the ratio between the circular polarization and the linear polarization. This parameter is independent of the absorption optical depth.
\label{fig2}}
\end{figure}


\begin{thebibliography}{}
\bibitem[Bersier et al.(2003)]{bersier03} Bersier, D., et al. 2003, \apj, 583, L63
\bibitem[Bhatt \& Bhattacharyya(2012)]{bhatt12} Bhatt, N., \& Bhattacharyya, S. 2012, \mnras, 420, 1706
\bibitem[Blandford \& McKee(1976)]{blandford76} Blandford, R. D., \& McKee, C. F. 1976, Physics of Fluids, 19, 1130
\bibitem[Burn(1966)]{burn66} Burn, B. J. 1966, \mnras, 133, 67
\bibitem[Chakraborty et al.(2015)]{cha15} Chakraborty, N., Pavlidou, V., \& Fields, B. D. 2015, \apj, 798, 16
\bibitem[Chang et al.(2014)]{chang14} Chang, Z., Lin, H.-N., Jiang, Y. 2014, \apj, 783, 30
\bibitem[Chatterjee et al.(2017)]{chatterjee17} Chatterjee, S., et al. 2017, Nature, 541, 58
\bibitem[Coburn \& Boggs(2003)]{coburn03} Coburn, W., \& Boggs, S. E. 2003, Nature, 423, 415
\bibitem[Covino et al.(1999)]{covino99} Covino, S., et al. 1999, \aap, 341, L1
\bibitem[Covino et al.(2015)]{covino15} Covino, S., et al. 2015, \aap, 578, A68
\bibitem[Cucchiara et al.(2011)]{cucchi11} Cucchiara, A., et al. 2011, \apj, 743, 154
\bibitem[de B\'{u}rca \& Shearer(2015)]{burca15} de B\'{u}rca, D. \& Shearer, A. 2015, \mnras, 450, 533
\bibitem[Eilek(1989)]{eilek89} Eilek, J. A. 1989, \aj, 98, 244
\bibitem[Fan et al.(2002)]{fan02} Fan, Y.-Z., Dai, Z.-G., Huang, Y.-F., \& Lu, T. 2002, ChJAA, 2, 449
\bibitem[Finkelstein et al.(2004)]{fink04} Finkelstein, A. M., Ipatov, A. V., Gnedin, Y. N., Ivanov, D .V., Kharinov, M. A., \& Rakhimov, I. A. 2004, Astronomy Letters, 30, 368
\bibitem[Frederiksen et al.(2010)]{fre10} Frederiksen, J. T., Haugb$\o$lle, T., Medvedev, M. V., \& Nordlund, \AA. 2010, \apj,
722, L114
\bibitem[Ghisellini \& Lazzati(1999)]{ghisellini99} Ghisellini, G., \& Lazzati, D. 1999, \mnras, 309, L7
\bibitem[Giannios \& Spitkovsky(2009)]{giannios09} Giannios, D., \& Spitkosky, A. 2009, \mnras, 400, 330
\bibitem[Gorbovskoy et al.(2016)]{gor16} Gorbovskoy, E. S., et al. 2016, \mnras, 455, 3312
\bibitem[G\"{o}tz et al.(2009)]{gotz09} G\"{o}tz, D., Laurent, P., Lebrun, F., Daigne, F., \& Bo\v{s}njak, \v{Z}. 2009, \apj, 695, L208
\bibitem[Granot(2003)]{granot03} Granot, J. 2003, \apj, 596, L17
\bibitem[Granot \& Taylor(2005)]{granot05} Granot, J. \& Taylor, J. B. 2005, \apj, 625, 263
\bibitem[Greiner et al.(2003)]{greiner03} Greiner, J., et al. 2003, Nature, 426, 157
\bibitem[Gruzinov \& Waxman(1999)]{gruzinov99} Gruzinov, A., \& Waxman, E. 1999, \apj, 511, 852
\bibitem[Gruzinov(1999)]{gruzinov99b} Gruzinov, A. 1999, \apj, 525, L29
\bibitem[Hjorth et al.(1999)]{hjorth99} Hjorth, J., et al. 1999, Science, 283, 2073
\bibitem[Ikejiri et al.(2011)]{ikejiri11} Ikejiri, Y., et al. 2011, PASJ, 63, 639
\bibitem[Jones \& O'Dell(1977)]{jones77} Jones, T. W., \& O'Dell, S. L. 1977, \apj, 214, 522
\bibitem[Kalemci et al.(2007)]{kalemci07} Kalemci, E., Boggs, S, E., Kouveliotou, C., Finger, M., \& Baring, M. G.
2007, ApJS, 169, 75
\bibitem[Kelner et al.(2013)]{kelner13} Kelner, S. R., Aharonian, F. A., \& Khangulyan, D. 2013, \apj, 774, 61
\bibitem[King et al.(2014)]{king14} King, O. G., et al. 2014, \mnras, 445, L114
\bibitem[Kobayashi(2000)]{kobayashi00} Kobayashi, S. 2000, \apj, 545, 807
\bibitem[Krawczynski et al. (2016)]{kra16} Krawczynski, H. S., et al. 2016, Astroparticle Physics, 75, 8
\bibitem[Laing(1980)]{laing80} Laing, R. A. 1980, \mnras, 193, 439
\bibitem[Laing(2002)]{laing02} Laing, R. A. 2002, \mnras, 329, 417
\bibitem[Lan et al.(2016)]{lan16} Lan, M.-X., Wu, X.-F., \& Dai, Z.-G. 2016, \apj, 816, 73
\bibitem[Landau \& Lifshitz(1971)]{landau71} Landau, L. D., \& Lifshitz, E. M. 1971, The Classical Theory of
Fields (Oxford: Pergamon)
\bibitem[Lazzati \& Begelman(2009)]{lazzati09} Lazzati, D., \& Begelman, M. C. 2009, \apj, 700, L141
\bibitem[Legg \& Westford(1968)]{legg68} Legg, M .P. C., \& Westford, K. C. 1968, \apj, 154, 499
\bibitem[Lemoine(2013)]{lemoine13} Lemoine, M. 2013, \mnras, 428, 845
\bibitem[Lorimer et al.(2007)]{lorimer07} Lorimer, D. R., Bailes, M., McLaughlin, M. A., Narkevic, D. J., \& Crawford, F. 2007, Science, 318, 777
\bibitem[Lyutikov(2006)]{lyutikov06} Lyutikov, M. 2006, \mnras, 369, L5
\bibitem[MacDonald \& Marscher(2017)]{macdonald17} MacDonald, N. R., \& Marscher, A. P. 2017, \apj, arXiv: 1611.09954
\bibitem[Macquart \& Melrose(2000a)]{macquart00a} Macquart, J.-P., \& Melrose, D. B. 2000a, Phys. Rev. E, 62, 4177
\bibitem[Macquart \& Melrose(2000b)]{macquart00b} Macquart, J.-P., \& Melrose, D. B. 2000b, \apj, 545, 798
\bibitem[Mao \& Wang(2011) hereafter MW11]{mao11} Mao, J., \& Wang, J. 2011, \apj, 731, 26
\bibitem[Mao \& Wang(2012)]{mao12} Mao, J., \& Wang, J. 2012, \apj, 748, 135
\bibitem[Mao \& Wang(2013)]{mao13} Mao, J., \& Wang, J. 2013, \apj, 776, 17
\bibitem[Martin \& Whittet(1990)]{martin90} Martin, P. G., \& Whittet, D. C. B. \apj, 357, 113
\bibitem[Matsumiya \& Ioka(2003)]{matsumiya03} Matsumiya, M., \& Ioka, K. 2003, \apj, 595, L25
\bibitem[Mead et al.(1990)]{mead90} Mead, A. R. G., Ballard, K. R., Brand, P. W. J. L., Hough, J. H., Brindle, C., \& Bailey, J. A. 1990,
A\&AS, 83, 183
\bibitem[McGlynn et al.(2007)]{mcglynn07} McGlynn, S. et al., 2007, \aap, 466, 895
\bibitem[Medvedev \& Loeb(1999)]{med99} Medvedev, M. V., \& Loeb, A. 1999, \apj, 526, 697
\bibitem[Medvedev(2000)]{medvedev00} Medvedev, M. V. 2000, \apj, 540, 704
\bibitem[Medvedev(2006)]{medvedev06} Medvedev, M. V. 2006, \apj, 637, 869
\bibitem[Melrose(1993a)]{melrose93a} Melrose, D. B., 1993a, J. Plasma Phys., 50, 267
\bibitem[Melrose(1993b)]{melrose93b} Melrose, D. B., 1993b, J. Plasma Phys., 50, 283
\bibitem[Melrose \& Macquart(1998)]{melrose98} Melrose, D. B., \& Macquart, J.-P. 1998, \apj, 505, 921
\bibitem[M\'{e}sz\'{a}ros \& Rees(1999)]{meszaros99} M\'{e}sz\'{a}ros, P., \& Rees, M. J. 1999, \mnras, 306, L39
\bibitem[Milosavljevi\'{c} et al.(2006)]{mi06} Milosavljevi\'{c} , M., Nakar, E., \& Spitkovsky, A. 2006, ApJ, 637, 765
\bibitem[Mundell et al.(2007)]{mundell07} Mundell, C. G., et al. 2007, Science, 315, 1822
\bibitem[Mundell et al.(2003)]{mundell13} Mundell, C. G., et al. 2013, Nature, 504, 119
\bibitem[Nava et al.(2016)]{nava15} Nava, L., Nakar, E., \& Piran, T. 2016, \mnras, 455, 1594
\bibitem[Orsi(2012)]{orsi12} Orsi, S. 2012, Nuclear Science Symposium and Medical Imaging Conference (NSS/MIC), p. 1880
\bibitem[Paul(2013)]{paul13} Paul, B. 2013, Int. J. Mod. Phys. D, 22, 1341009
\bibitem[Petroff et al.(2015)]{petroff15} Petroff, E., et al. 2015, \mnras, 447, 246
\bibitem[Piran(1999)]{piran99} Piran, T. 1999, Physics Report, 314, 575
\bibitem[Prosekin et al.(2016)]{pro16} Prosekin, A. Y., Kelner, S. R., \& Aharonian, F. A. 2016, Phys. Rev. D. 94, 063010
\bibitem[Rol et al.(2000)]{rol00} Rol, E., et al. 2000, \apj, 544, 707
\bibitem[Rol et al.(2003)]{rol03} Rol, E., et al. 2003, \aap, 405, L23
\bibitem[Rossi et al.(2004)]{rossi04} Rossi, E. M., Lazzati, D., Salmonson, J. D., \& Ghisellini, G. 2004, \mnras, 354, 86
\bibitem[Rybicki \& Lightman(1979)]{rybicki79} Rybicki, G., \& Lightman, A. P., 1979, Radiative Processes in Astrophysics.,
Wiely Interscience, New York
\bibitem[Sagiv etal.(2004)]{sagiv04} Sagiv, A., Waxman, E. \& Loeb, A. 2004, \apj, 615, 366
\bibitem[Sari \& Piran(1999)]{sari99} Sari, R., \& Piran, T. 1999, \apj, 520, 641
\bibitem[Sazonov \& Tsytovich(1968)]{sazonov68} Sazonov, V. N., \& Tsytovich, V. N. 1968, Radiofizika, 11, 1287
\bibitem[Sazonov(1969a)]{sazonov69a} Sazonov, V. N. 1969a, Soviet Astron., 13, 396
\bibitem[Sazonov(1969b)]{sazonov69b} Sazonov, V. N. 1969b, Soviet Physics JETP, 29, 578
\bibitem[Schekochihin \& Cowley (2007)]{schekochihin07} Schekochihin, A. A., \& Cowley, S. T. 2007, in Magnetohydrodynamics:
Historical Evolution and Trends, ed. S. Molokov, R. Moreau, \& H. K. Moffatt (Berlin: Springer), 85
\bibitem[Schekochihin et al.(2004)]{sch04} Schekochihin, A. A., Cowley, S. C., Tayler, S. F., Marson, J. A., \& McWilliams, J. C. 2004, \apj, 612, 276
\bibitem[Serkowski et al.(1975)]{serkowski75} Serkowski, K., Matheson, D. L., \& Ford, V. L. 1975, \apj, 196, 261
\bibitem[She \& Leveque(1994)]{she94} She, Z.-S., \& Leveque, E. 1994, Phys. Rev. Lett., 72, 336
\bibitem[She \& Waymire(1995)]{she95} She, Z.-S., \& Waymire, E. C. 1995, Phys. Rev. Lett., 74, 262
\bibitem[Simonetti et al.(1984)]{simonetti84} Simonetti, J. H., Cordes, J. M., \& Spangler, S. R. 1984, \apj, 284, 126
\bibitem[Sironi \& Spitkovsky(2009)]{sirony09} Sironi, L., \& Spitkovsky, A., 2009, \apj, 707, L92
\bibitem[Sokoloff et al.(1998)]{sokoloff98} Sokoloff, D. D., Bykov, A. A., Shukurov, A., Berkhuijsen, E. M., Beck, R., \& Poezd, A. D. 1998, \mnras, 299, 189
\bibitem[Stawarz \& Petrosian(2008)]{stawarz08} Stawarz, L., \& Petrosian, V. 2008, \apj, 681, 1725
\bibitem[Steele et al.(2009)]{steele09} Steele, I. A., Mundell, C. G., Smith, R. J., Kobayashi, S., \& Guidorzi, C. 2009, Nature, 462, 767
\bibitem[Taylor et al.(2005)]{taylor05} Taylor, G. B., Momjian, E., Pihlstr\"{o}m, Y., Ghosh, T., \& Salter, C. 2005, \apj, 622, 986
\bibitem[Teraki \& Takahara(2013)]{teraki13} Teraki, Y., \& Takahara, F. 2013, \apj, 787, 28
\bibitem[Toma et al.(2008)]{toma08} Toma, K., Ioka, K., \& Nakamura, T. 2008, ApJ, 673, L123
\bibitem[Tribble(1991)]{tribble91} Tribble, P. C. 1991, \mnras, 250, 726
\bibitem[Uehara et al.(2012)]{uehara12} Uehara, T., et al. 2012, \apj, 752, L6
\bibitem[Wang et al.(2015)]{wang15} Wang, S., et al. 2015, 7th symposium on large TPCs for low-energy rare event detection, Paris, France, arXiv: 1503.03772
\bibitem[Westford(1959)]{westford59} Westford, K. C. \apj, 130, 241
\bibitem[Wiersema et al.(2012)]{wier12} Wiersema, K., et al. 2012, \mnras, 426, 2
\bibitem[Wiersema et al.(2014)]{wier14} Wiersema, K., et al. 2014, Nature, 509, 201
\bibitem[Wijers et al.(1999)]{wijers99} Wijers, R. A. M., et al. 1999, \apj, 523, L33
\bibitem[Workman et al.(2008)]{workman08} Workman, J. C., Morsony, B. J., Lazatti, D., Medvedev, M. V. 2008, \mnras, 386, 199
\bibitem[Yatsu et al.(2014)]{yatsu14} Yatsu, Y., et al. 2014, Proc. SPIE, 9144, 91440L
\bibitem[Yonetoku et al.(2011)]{yonetoku11} Yonetoku, D., et al. 2011, \apj, 743, L30
\bibitem[Yonetoku et al.(2012)]{yonetoku12} Yonetoku, D., et al. 2012, \apj, 758, L1
\bibitem[Zhang et al.(2003)]{zhang03} Zhang, B., Kobayashi, S., \& M\'{e}sz\'{a}ros, P. 2003, \apj, 595, 950
\bibitem[Zrake \& MacFadyen(2012)]{zrake13} Zrake, J., \& MacFadyen, A. I. 2013, \apj, 763, L12
\end{thebibliography}
\end{document}